\documentclass[%
 reprint,
superscriptaddress,
 amsmath,amssymb,
 aps,
]{revtex4-1}

\usepackage{graphicx}
\usepackage{dcolumn}
\usepackage{bm}
\usepackage{gensymb}	
\usepackage{epsfig}	
\usepackage{multirow} 

\begin{document}

\preprint{APS/123-QED}

\title{Infrared study of carrier scattering mechanism in ion-gated graphene}

\author{Kwangnam Yu}
\affiliation{Department of Physics, University of Seoul, Seoul 130-743, Republic of Korea}

\author{Jiwon Jeon}
\affiliation{Department of Physics, University of Seoul, Seoul 130-743, Republic of Korea}

\author{Jiho Kim}
\affiliation{Department of Physics, University of Seoul, Seoul 130-743, Republic of Korea}

\author{Chang Won Oh}
\affiliation{Department of Physics, University of Seoul, Seoul 130-743, Republic of Korea}

\author{Yongseok Yoon}
\affiliation{Department of Physics, University of Seoul, Seoul 130-743, Republic of Korea}

\author{Beom Joon Kim}
\affiliation{SKKU Advanced Institute of Nanotechnology (SAINT), Sungkyunkwan University, Suwon 440-746, Korea}

\author{Jeong Ho Cho}
\affiliation{Department of Chemical and Biomolecular Engineering, Yonsei University, Seoul 03722, Republic of Korea}

\author{E. J. Choi}
\email[Corresponding author: ]{echoi@uos.ac.kr}
\affiliation{Department of Physics, University of Seoul, Seoul 130-743, Republic of Korea}

\date{\today}

\begin{abstract}
We performed infrared transmission experiment on ion-gel gated graphene and measured carrier scattering rate $\gamma$ as 
function of carrier density $n$ over wide range up to $n=2\times10^{13}$~cm$^{-2}$. 
The $\gamma$ exhibits a rapid decreases along with the gating followed by persistent increases on further carrier doping. 
This behavior of $\gamma(n)$ demonstrates that carrier is scattered dominantly by the two scattering mechanisms, 
namely, charged impurity (CI) scattering and short-range disorder (SR) scattering, with additional minor scattering from substrate phonon (SPP).
We can determine the absolute strengths of all the scattering channels by fitting the $\gamma(n)$ data and 
unveils the complete $n$-dependent map of the scattering mechanisms $\gamma(n)=\gamma_{\rm{CI}}(n)+\gamma_{\rm{SR}}(n)+\gamma_{\rm{SPP}}(n)$.
The $\gamma_{\rm{CI}}(n)$ and $\gamma_{\rm{SR}}(n)$ are larger than those of SiO$_{2}$-gated graphene by 1.8 times,
which elucidates the dual role of the ion-gel layer as a CI-scatterer and simultaneously a SR-scatterer to graphene.
Additionally we show that freezing of IG at low-$T$ ($\sim200$~K) does not cause any change to the carrier scattering. 
\end{abstract}


\maketitle

Graphene holds great promise for application due to excellent transport, optical, and mechanical properties.
\cite{novoselov2005two, li2008dirac, lee2008measurement, neto2009electronic, ju2011graphene, yan2012tunable}
In particular the massless charge carrier that obeys the Dirac electrodynamics can support a high-mobility two-dimensional electrical conduction, 
the property which can be utilized for high-speed electronic device.\cite{zhang2005experimental, bolotin2008ultrahigh}
However for the large-scale CVD-grown graphene that is suitable for practical applications, 
the carrier mobility is generally inferior compared with that of its exfoliated counterpart. 
This fact shows that in the CVD graphene the carrier is scattered more oftenly, which again implies that there are more scattering sources.
\cite{pirkle2011effect, vlassiouk2011electrical, wang2011electrochemical}
To improve the carrier mobility of the large-scale graphene, complete scattering mechanism should be elucidated by theoretical and experimental studies.  

In graphene the carrier is scattered by various scattering sources such as charged-impurity (CI) and short-ranged disorder (SR), 
acoustic/optical phonon of graphene, and the substrate.\cite{hwang2007carrier, hwang2008acoustic, sarma2011electronic, hwang2013surface}
When the electrical resistivity or carrier mobility is experimentally measured, the scatterings caused by those sources are all added up.
Therefore it is difficult to find out which scattering(s) is the major one that is dominantly degrading the mobility.
However, importantly, the interaction between the carrier with the scattering sources is predicted to be dependent on the carrier density ($n$) of graphene: 
When $n$ is increased the CI-scattering decreases while the SR-scattering increases.
Specifically their scattering rate ($\gamma$) change with $n$ as, according to theoretical calculation, 
$\gamma_{\rm{CI}} = A/\sqrt{n}$ and $\gamma_{\rm{SR}} = B\sqrt{n}$.\cite{hwang2007carrier, sarma2011electronic}
As for optical phonon of graphene, the scattering rate is independent of $n$, $\gamma_{\rm{OP}} = \gamma_{0}$.\cite{hwang2008acoustic}
Such distinct $n$-dependent scattering rates characteristic of the scattering sources can be used to determine them separately. 
That is, if one can measure the experimental $\gamma$ while controlling $n$ over wide range   
and compare the result $\gamma(n)$ with the $n$-dependent $\gamma$'s of all the scattering sources 
it may be possible to determine how strongly each of those scatterings contributes to the total scattering.
The complete scattering mechanisms elucidated in this manner can provide the fundamental knowledge on the engineering effort toward 
low-scattering graphene with ultra-high carrier mobility.

Here we perform infrared transmission measurement of ion-gel gated graphene.
In the ion-gel gating device the electronic double layers (EDL), i.e, ion-layer and graphene-layer, 
are separated by nano-scale air-gap, which builds a parallel-plate capacitor with very large charge capacitance $C\sim10~\mu$F.\cite{lee2007ion}
As result, carrier density as high as $n\sim 1\times10^{14}$~cm$^{-2}$ is induced in graphene at low-voltage (only a few volt) gating.
The high-$n$ doping enables the emergence of the novel effects in 2d materials and oxide compounds such as 
metal-insulator transition,\cite{shimotani2007insulator,jeong2013suppression} 
gate-tunable 2d ferromagnetism,\cite{moetakef2012carrier, deng2018gate} and the superconducting transition.\cite{ye2010liquid,taniguchi2012electric, costanzo2016gate}
The ion gating is also suitable for our scattering mechanism study because the $\gamma$ can be measured over wide range of $n$. 
We remark that in the EDL, the carrier can be scattered by the ion-layer in addition to the impurity and phonon scattering sources mentioned above,
because the ion-layer lies only a few nm away from graphene.
The carrier scattering due to the ion-layer is not well understood at this point not only for graphene-EDL but also for many other ion-gated devices in general.


Large scale CVD graphene ($1~\rm{cm}\times1~\rm{cm}$) was synthesized and transferred on SiO$_{2}$(300~nm thick)/p-Si substrate. 
\cite{kim2009large,li2009transfer, bae2010roll, lee2010wafer}
Here IR-transparent un-doped p-Si substrate was used for the transmission measurement.
The thickness characterization of the sample was performed and described in previous publications.\cite{lee2016strong, yu2016infrared} 
The sample was annealed at high temperature $T=520$~K for 2 hours in vacuum ($P=400$~mTorr).  
To perform the ion-gel gating on the graphene, mixture of [EMIM][TFSI] ionic liquid, PS-PMMA-PS triblock copolymer, 
and ethyl acetate solvent (weight ratio = 0.1 : 0.9 : 9) was prepared and spin-coated on the sample.\cite{kim2010high} 
The spin-coating was applied in less than 15 min after the sample was taken out of the vacuum to minimize the exposure to air. 
Infrared transmittance was measured over the frequency range from $\omega=50$~cm$^{-1}$ to $250$~cm$^{-1}$ using a Fourier Transform Infrared (FTIR) spectrometer 
and a 4.2~K-cooled bolometric detector. The incident IR beam size was 3~mm in diameter.



\begin{figure}[t]
\centerline
{\includegraphics[width=0.44\textwidth]{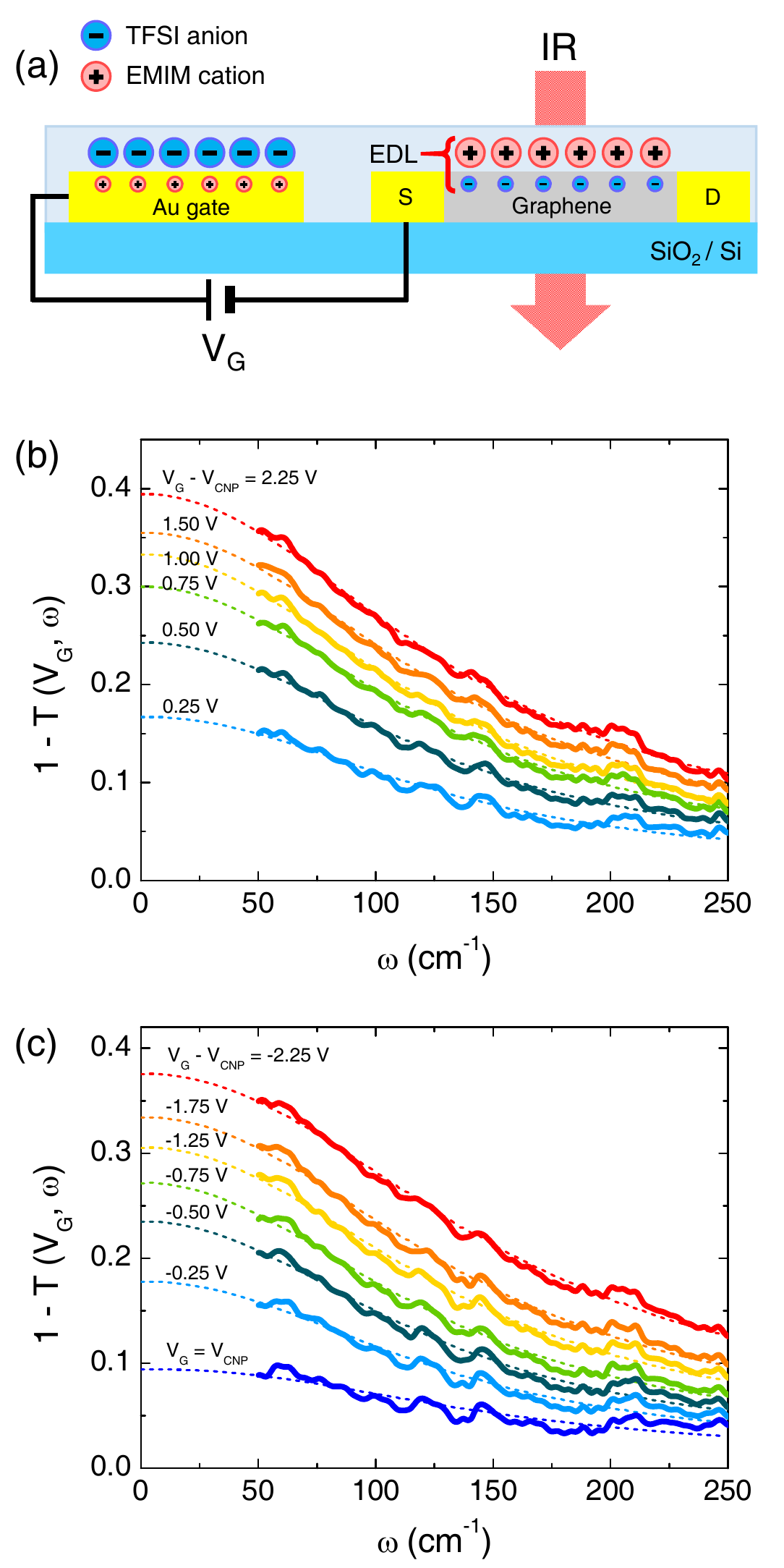}}
\caption{
(a) Schematic diagram of infrared transmission measurement on ion-gel gated graphene device. 
As gate-voltage $V_{\rm{G}}$ is applied, the ionic charge and graphene carrier are induced in the EDL (electronic double layer) device.
(b) $1-T(\omega)$ for the electron-doping regime and (c) hole-doping regime. 
$T(\omega)$ stands for the normalized transmission data (see text). 
}
\label{fig:1}
\end{figure}

Figure~\ref{fig:1}(a) shows schematic diagram of the ionic gating and the infrared transmission measurement. 
When the gate voltage $V_{\rm{G}}$ is applied, ionic charge and Dirac carrier are induced in the ion-gel (IG) and graphene respectively,
creating the electronic double layer (EDL). 
We measured infrared transmittance through the EDL at fixed $V_{\rm{G}}$ and then normalized it 
by the reference transmittance measured on bare-IG/substrate (no graphene), which we label as $T(\omega)$ hereafter. 
Figure~\ref{fig:1}(b) and (c) show $1-T(\omega)$ for the hole-doping regime ($V_{\rm{G}}-V_{\rm{CNP}}<0$) and electron-doping regime ($V_{\rm{G}}-V_{\rm{CNP}}>0$) respectively, 
where the transport $I-V$ measurement showed $V_{\rm{CNP}}=0.1$~V for the same sample.
We remark that  $1-T(\omega)$ represents the Drude absorption due to the graphene carrier.
As $V_{\rm{G}}$ is increased, the carrier in graphene increases and the Drude absorption becomes stronger.
For quantitative analysis we performed rigorous fitting of $1-T(\omega)$ using the Drude conductivity model for graphene
\begin{equation}
\sigma(\omega) = \frac{\sigma_{\rm{DC}}}{1+i\cdot{\omega}/{\gamma}}
\label{eq:1}
\end{equation}
where the two fitting parameters $\sigma_{\rm{DC}}$ and $\gamma$ represent DC-conductivity and scattering rate of the carrier respectively.
For the multilayer optical analysis we used the RefFit program.\cite{kuzmenko2005kramers}
The fitting curves (dashed) show good agreement with data. 


\begin{figure}[t]
\centerline
{\includegraphics[width=0.44\textwidth]{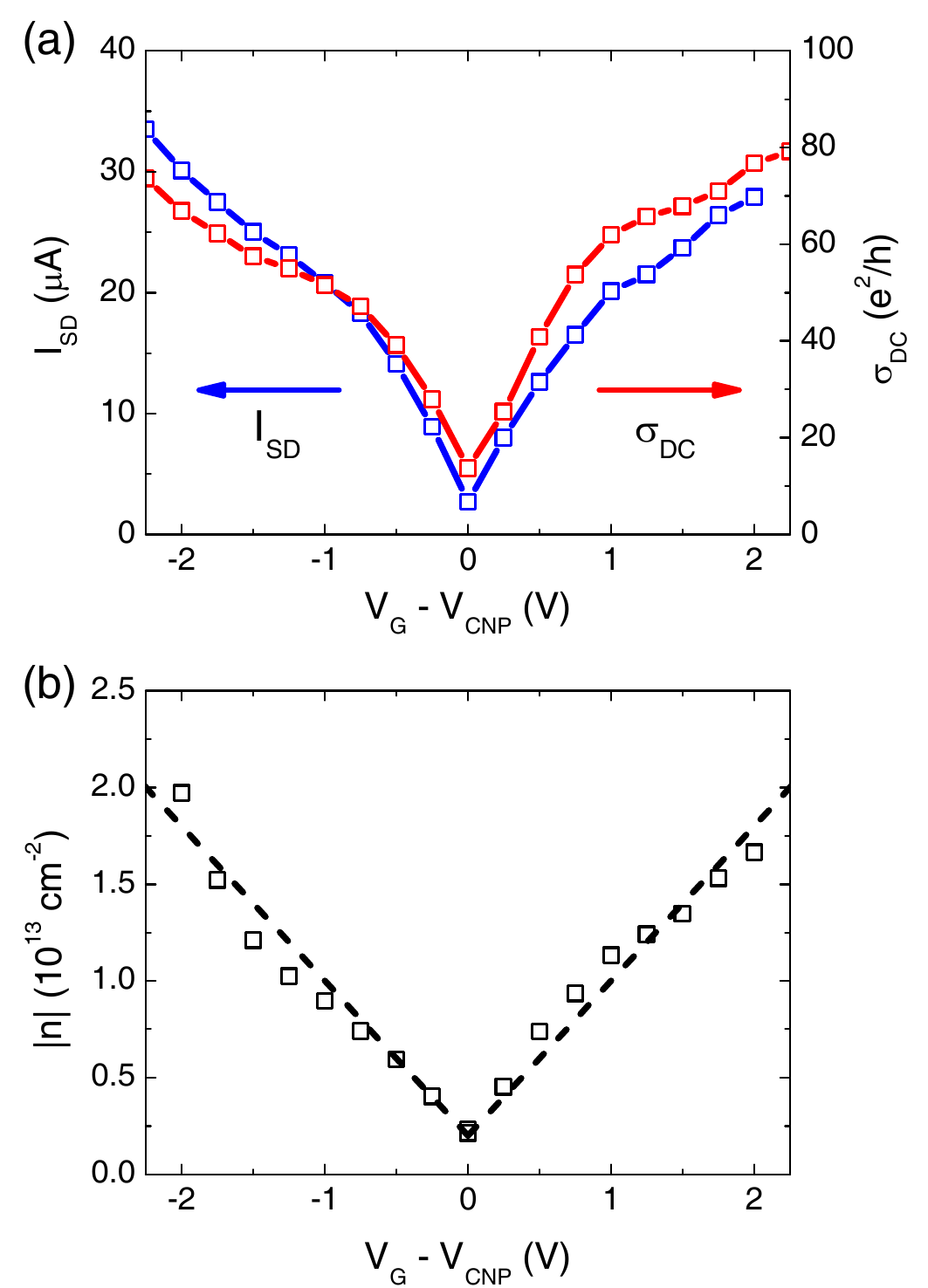}}
\caption{
(a) DC-conductivity $\sigma_{\rm{DC}}$ and the dc-current $I_{SD}$ are compared. 
They are determined from the IR transmission data and the transport $I$-$V$ data, respectively. 
(b) Carrier density $|n|$ measured from the Drude absorption of graphene. 
}
\label{fig:2}
\end{figure}

Figure~\ref{fig:2}(a) shows the fitting result $\sigma_{\rm{DC}}$ and $\gamma$. 
$\sigma_{\rm{DC}}$ increases as $V_{\rm{G}}$ is applied. 
This behavior is in reasonable agreement with the $I$-$V$ curve demonstrating the consistency between the infrared and transport measurements.
Before we show $\gamma$, we first calculate carrier density $n$ from the measured $\sigma_{\rm{DC}}$ and $\gamma$ using the relation 
$\sigma_{\rm{DC}}=(v_{\rm{F}}e^{2}/\sqrt{\pi}\hbar)(\sqrt{|n|}/\gamma)$.\cite{ando2006screening, nomura2007quantum, hwang2007carrier}
Here Fermi velocity $v_{\rm{F}}=1.1\times10^{6}$~m/s is taken.\cite{zhang2005experimental}
Figure~\ref{fig:2}(b) shows that $n$ increases linearly with $V_{\rm{G}}-V_{\rm{CNP}}$.
The carrier density reaches $n=1.5\times10^{13}$~cm$^{-2}$ at $V_{\rm{G}}-V_{\rm{CNP}}=\pm1.75$~V demonstrating the high-density carrier doping at low-voltage. 
Previous transport measurements showed that $n\sim10^{14}$~cm$^{-2}$ can be reached for small devices, $\sim10~\mu m$ in length.\cite{ye2011accessing, efetov2010controlling}
When we compare the $n$ with the charge $Q=CV$ of capacitor ($Q=enA$, $A$= gated area) the capacitance of our device is found to be $C/A=1.22~\mu$F~cm$^{-2}$.

\begin{figure}[t]
\centerline
{\includegraphics[width=0.44\textwidth]{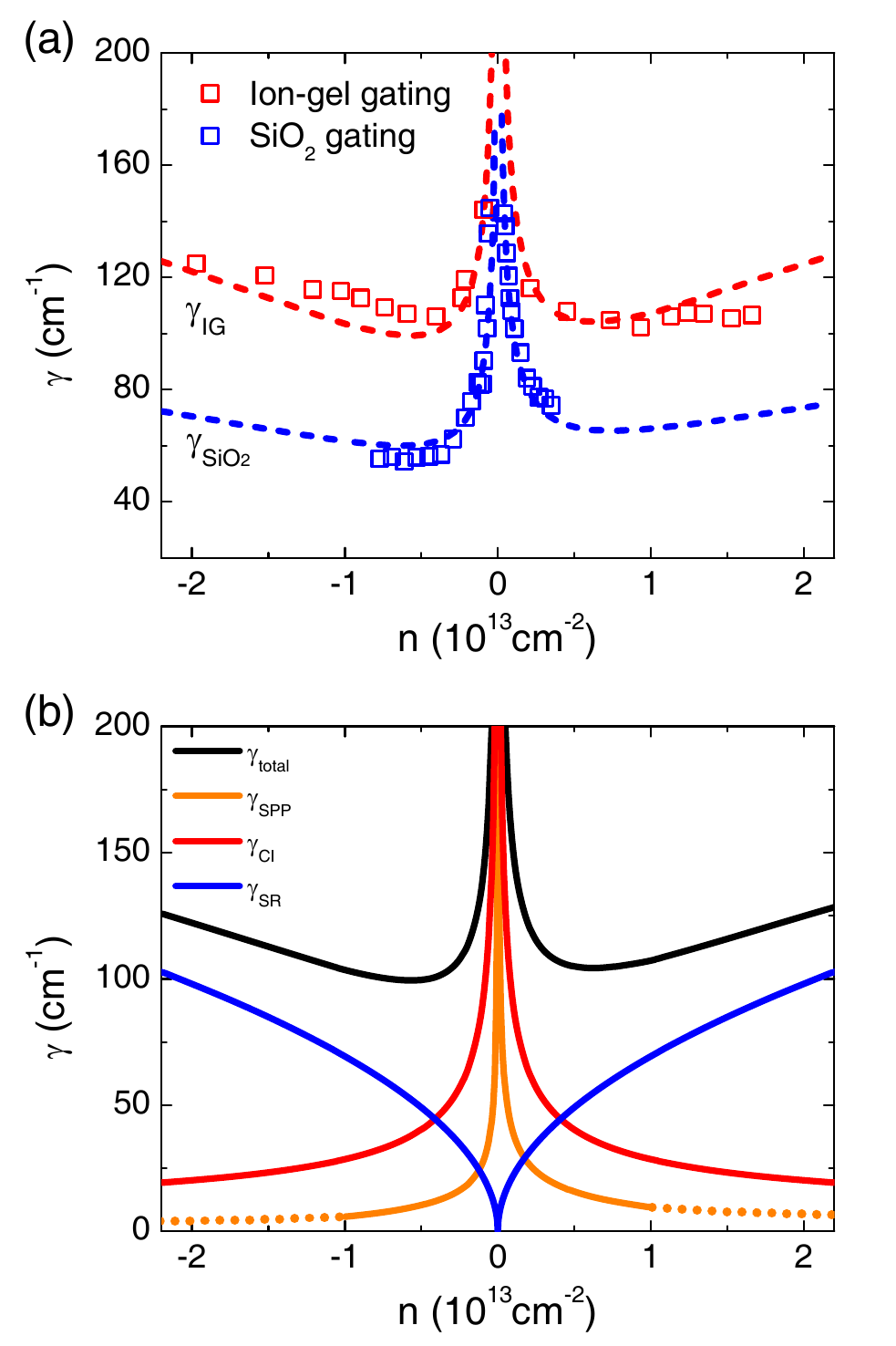}}
\caption{
(a) Carrier scattering rate $\gamma(n)$ determined from the Drude fitting. 
For comparison the $\gamma(n)$ measured for back-gated graphene/SiO$_{2}$/Si device is shown together.\cite{yu2016infrared}. The dashed curves show the fitting results.
(b) The three scattering rates are shown as function of the carrier density: 
$\gamma_{\rm{CI}}$ for charged impurity, $\gamma_{\rm{SR}}$ for short-range disorder, $\gamma_{\rm{SPP}}$ for surface polar phonon of SiO$_{2}$.
The total scattering $\gamma=\gamma_{\rm{CI}}+\gamma_{\rm{SR}}+\gamma_{\rm{SPP}}$ is shown by the black curve.
}
\label{fig:3}
\end{figure}

Given the $n$ for all the $V_{\rm{G}}$'s, we can plot $\gamma$ as function of $n$.
Figure~\ref{fig:3}(a) shows that $\gamma$ decreases rapidly with the $n$-increase at low-doping regime which is followed by an upturn upon further doping. 
This behavior can be explained if we assume the carrier scattering arises due to three scattering sources:
charged-impurity (CI), short-range disorder (SR), and the phonons of graphene and substrate.
\begin{equation}
\gamma=\gamma_{\rm{CI}}+\gamma_{\rm{SR}}+\gamma_{\rm{SPP}}
\label{eq:2}
\end{equation}  
The scattering due to CI and SR depend on $n$ as $\gamma_{\rm{CI}}=A/\sqrt{n}$ and $\gamma_{\rm{SR}}=B\sqrt{n}$, respectively. \cite{sarma2011electronic}
The phonon scattering was explicitly measured for SiO$_{2}$-gated graphene, 
which showed that the surface polar phonon (SPP) of SiO$_{2}$-substrate is dominant over the other phonons.\cite{yu2016infrared}
We fit the $\gamma(n)$ data using $A$ and $B$ as fitting variables. 
For $\gamma_{\rm{SPP}}(n)$ the measured data in Ref.~\cite{yu2016infrared} is inserted into Eq.~(\ref{eq:2}).
The fit shows reasonable agreement with data.
In Figure~\ref{fig:3}(b) we show the fitting curves $\gamma_{\rm{CI}}$, $\gamma_{\rm{SR}}$, and $\gamma_{\rm{SPP}}$.
They show that at very low-doping near $n=0$,  $\gamma_{\rm{CI}}$ is much larger than the others, revealling  
that the rapid decrease of the total scattering rate $\gamma(n)$ arises due to the dominant role of $\gamma_{\rm{CI}}$. 
As $n$ increases, $\gamma_{\rm{CI}}$ decays quickly and instead $\gamma_{\rm{SR}}$ becomes the major scattering rate.
The $\gamma_{\rm{SR}}=B\sqrt{n}$ behavior was theoretically predicted \cite{sarma2011electronic} and also was used previously to explain SiO$_{2}$-gated scattering result. \cite{yu2016infrared}
There, however, the doping range was limited to only $n<8\times10^{12}$~cm$^{-2}$ (see the blue curve of Fig.~\ref{fig:3}(a)). 
Therefore the validity of the $\sqrt{n}$-dependent $\gamma_{\rm{SR}}$ was seriously questioned.
In this IG-gated experiment the doping range is largely extended compared with the SiO$_{2}$-gating.
Here we observe that $\gamma(n)$ increases persistently over the wide range of $n$ up to $2\times10^{13}$~cm$^{-2}$,
unambiguously demonstrating the $\gamma_{\rm{SR}}=B\sqrt{n}$ of the SR-scattering.
Given the $n$-dependent $\gamma_{\rm{CI}}$ and $\gamma_{\rm{SR}}$ proved over the wide-range, 
it is clear that CI, SR, and SPP are the major scattering sources of graphene.

From the fit we have $A=9.0\times10^{7}$~cm$^{-2}$ and $B=2.2\times10^{-5}$. 
They are larger than those of the SiO$_{2}$-gated graphene ($A=5.0\times10^{7}$~cm$^{-2}$, $B=1.2\times10^{-5}$) 
by the ratios $A_{\rm{IG}}/A_{\rm{SiO_{2}}}=1.8$ and $B_{\rm{IG}}/B_{\rm{SiO_{2}}}=1.8$ that shows the scatterings are stronger in the IG-gated graphene.
Note that in the EDL-device carrier can be scattered by the IG-layer in addition to the CI, SR, and SPP scatterers. 
Specifically the IG molecules [EMIM]$^{+}$ and [TFSI]$^{-}$ can exert long-range Coulomb scattering on the carriers.
Because the IG-layer is in close proximity with graphene, this Coulomb scattering can be of significant strength. 
The measured $A$-ratio shows that the IG-driven Coulomb scattering is comparable to the CI-driven Coulomb scattering (0.8 : 1). 
As for $\gamma_{\rm{SR}}$, the short-range scattering arises from topological lattice disorder of graphene such as single atomic vacancy and 
few-atomic vacancy cluster like pentagons/heptagons.\cite{ebbesen1995topological} 
They are described as the hard-sphere potential in scattering theory and, due to their short-ranged interaction, create the inter-valley scattering in graphene.
The increase of the $B$-coefficient shows that SR-scattering is stronger in the EDL-device. 
This result is however difficult to explain: note that the IG-molecules consist of long chemical chains, by-far larger than the atomic defects.
Moreover, the carriers do not scatter with the hard-sphere potential of IG-molecule because the IG-layer is lying away from the conducting path, graphene. 
Therefore it is not likely that the IG-layer accounts for the $\gamma_{\rm{SR}}$-increase.
Alternatively one may postulate that IG molecules chemically react with graphene, creating the atomic or cluster defect. 
However it is generally agreed that graphene is stable against ion-gel unlike some oxide materials that are believed to chemically react. 
At this point the mechanism for the $\gamma_{\rm{SR}}$-increase in the IG-gated graphene is not clear.
We think that further study is needed to unveil the non-trivial interplay between IG and graphene. 
  
We remark that for Si-gated there was another scattering  $\gamma_{\rm{AD}}$ due to air/water molecules adsorbed on graphene 
because the graphene was exposed to air during the gating.\cite{yu2016infrared}
The $\gamma_{\rm{AD}}(n)$ is asymmetric for the hole- and electron- doping unlike other $\gamma$'s. 
For IG-gated graphene $\gamma(n)$ can be fit without need of the asymmetric $\gamma_{\rm{AD}}$ term. 
This appears to demonstrate that in the IG-gated graphene, the IG-layer prevents the air from adhering to graphene.
 
\begin{figure}[t]
\centerline
{\includegraphics[width=0.44\textwidth]{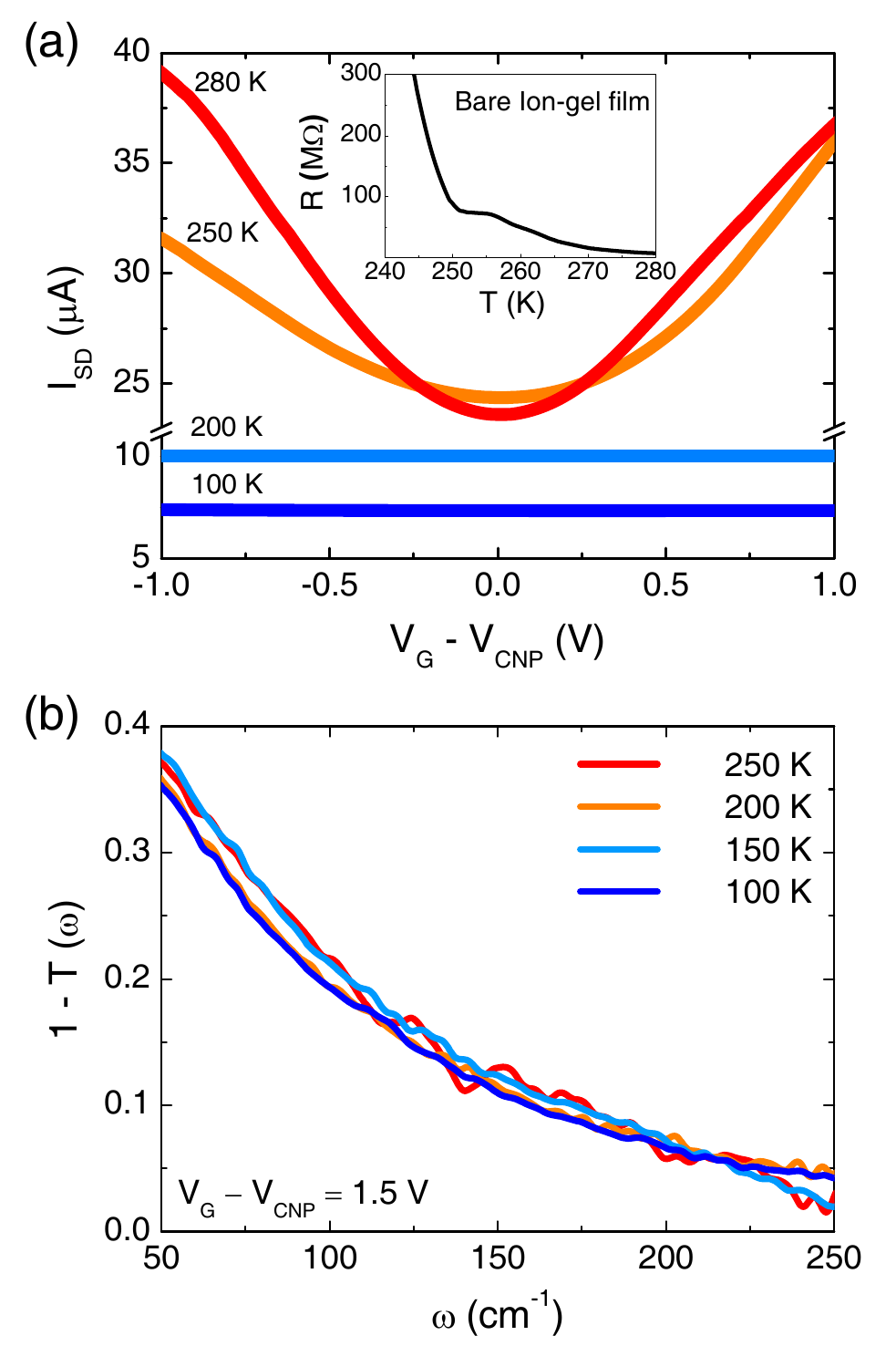}}
\caption{
(a) Temperature dependent $I-V$ curve measured for IG-gated graphene. Inset shows temperature dependent DC-resistance $R(T)$ of bare IG thin film. 
(b) Far-infrared Drude absorption of graphene for $100~{\rm{K}}<T<300~{\rm{K}}$ temperature range.  
The gate voltage $V_{\rm{G}}-V_{\rm{CNP}}=1.5$~V was applied at $T=300$~K and maintained during the low-$T$ measurement. 
}
\label{fig:4}
\end{figure}

For complete understanding of the effect of IG on scattering, we further measure $\gamma$ at low-$T$.
When temperature decreases, IG freezes at a certain freezing-$T$. 
Whether the IG-freezing brings any additional scattering or not is an unanswered question.  
To investigate this issue is of significant importance when noting that novel electronic phases such as 
ferromagnetism and superconductivity can emerge in 2d-materials by the IG-gating. 
Here we measured $T$-dependent transport and infrared properties of the bare IG and IG/graphene.
Figure~\ref{fig:4}(a) shows that in the $I-V$ curve, the current stops responding to $V_{\rm{G}}$ at $T<200$~K demonstrating 
that IG molecules become immobile due to the freezing.  
The inset shows DC-resistance $R(T)$ of the bare IG thin film (no graphene) deposited on SiO$_{2}$/Si substrate. 
Here the IG film was $20~\mu m$-thick and $5~\rm{mm} \times 5~\rm{mm}$ in area, respectively. 
Au-contact was made on the edges of the IG for two-probe resistance measurement. 
The $R(T)$ increases strongly as $T$ decreases, exceeding the instrumental limit at $T<240$~K, consistent with the freezing of IG at $T\sim200$~K. 
Given this transport result, we applied $V_{\rm{G}}$ at $T=300$~K and measured the Drude absorption while cooling down the EDL-device to 100~K. 
Figure~\ref{fig:4}(b) shows that the Drude peak shows no change as $T$ passes through the freezing $T$
in the strength and width (i.e, carrier density and scattering rate), which reveals conclusively that the IG-freezing does not perturb the carrier conduction in graphene.
We think that this results applies to not only graphene but also many other channel materials of EDL-devices.   

To summarize we carried out infrared transmission measurement and measured carrier scattering rate $\gamma$ of ion-gel gated graphene to investigate carrier scattering mechanisms.
As the gating is applied, $\gamma$ decreases rapidly in the initial stage and then increases persistently on further carrier doping. 
This non-trivial gate-dependent change shows that carrier is scattered by the three scattering sources, 
namely, charged impurity (CI), short-range impurity(SR), and substrate phonon (SPP) as $\gamma(n)=\gamma_{\rm{CI}}(n)+\gamma_{\rm{SR}}(n)+\gamma_{\rm{SPP}}(n)$. 
Our IG-gating measurement demonstrates unambiguously that the CI-scattering and SR-scattering exhibit 
the $n$-dependent behaviors $\gamma_{\rm{CI}}=A/\sqrt{n}$ and $\gamma_{\rm{SR}}=B\sqrt{n}$, 
which are verified by the data-fit over the wide-range of $n$ up to $2\times10^{13}$~cm$^{-2}$. 
The CI-scattering and SR-scattering are the dominant scatterings at the low-doping and high-doping regimes of graphene, respectively.
Furthermore we found that $A$ and $B$ are definitely larger than those of SiO$_{2}$-gated graphene,
which shows that the IG-layer acts as CI-scatterer and simultaneously SR-scatterer to graphene.
Additionally $\gamma$ measured from room-T down to $T=100$~K shows no temperature-dependent 
change demonstrating that the low-$T$ freezing of IG does not affect the carrier scattering. 

This work research was supported by the 2016 Research Fund of the University of Seoul.

%

\end{document}